\documentclass{article}

\usepackage[preprint]{neurips_2025}

\usepackage[utf8]{inputenc} 
\usepackage[T1]{fontenc}    
\usepackage{hyperref}       
\usepackage{url}            
\usepackage{booktabs}       
\usepackage{amsfonts}       
\usepackage{nicefrac}       
\usepackage{microtype}      
\usepackage{xcolor}         
\usepackage{amsmath}
\usepackage{array}
\usepackage{graphicx}
\usepackage{wrapfig}
\usepackage{cleveref}

\newcommand{\startappendix}{%
    \setcounter{figure}{0}%
    \setcounter{table}{0}%
    \renewcommand{\thefigure}{S\arabic{figure}}%
    \renewcommand{\thetable}{S\arabic{table}}%
    \crefname{figure}{Figure S}{Figures S}%
    \crefname{table}{Table S}{Tables S}%
    \renewcommand{\thesection}{\Alph{section}}       
    \renewcommand{\thesubsection}{\thesection.\arabic{subsection}}  

}

\title{Capturing Aperiodic Temporal Dynamics of EEG Signals through Stochastic Fluctuation Modeling}

\author{
  Yuhao Sun$\textsuperscript{\rm $1,2$}$\thanks{Equal contribution}, Zhiyuan Ma$\textsuperscript{\rm $1,2,3$}$\footnotemark[1], Xinke Shen$\textsuperscript{\rm $4$}$, Jinhao Li$\textsuperscript{\rm $1,5$}$, Guan Wang$\textsuperscript{\rm $6$}$, Sen Song$\textsuperscript{\rm $1,2$}$\thanks{Corresponding author}\\
  $\textsuperscript{\rm $1$}$ Laboratory of Brain and Intelligence, Tsinghua University\\
  $\textsuperscript{\rm $2$}$ School of Biomedical Engineering, Tsinghua University\\
  $\textsuperscript{\rm $3$}$ College of Computer Science and Technology, Zhejiang University\\
  $\textsuperscript{\rm $4$}$ The Department of Biomedical Engineering, Southern University of Science and Technology\\
  $\textsuperscript{\rm $5$}$School of Basic Medical Sciences, Tsinghua University\\
  $\textsuperscript{\rm $6$}$ Sapient Intelligence\\
  \texttt{syh18@mails.tsinghua.edu.cn} \\
  \texttt{songsen@tsinghua.edu.cn}
}

\begin{document}

\maketitle

\begin{abstract}

Electrophysiological brain signals, such as electroencephalography (EEG), exhibit both periodic and aperiodic components, with the latter often modeled as 1/f noise and considered critical to cognitive and neurological processes.
Although various theoretical frameworks have been proposed to account for aperiodic activity, its scale-invariant and long-range temporal dependency remain insufficiently explained.
Drawing on neural fluctuation theory, we propose a novel framework that parameterizes intrinsic stochastic neural fluctuations to account for aperiodic dynamics.
Within this framework, we introduce two key parameters—self-similarity and scale factor—to characterize these fluctuations.
Our findings reveal that EEG fluctuations exhibit self-similar and non-stable statistical properties, challenging the assumptions of conventional stochastic models in neural dynamical modeling.
Furthermore, the proposed parameters enable the reconstruction of EEG-like signals that faithfully replicate the aperiodic spectrum, including the characteristic 1/f spectral profile, and long range dependency.
By linking structured neural fluctuations to empirically observed aperiodic EEG activity, this work offers deeper mechanistic insights into brain dynamics, resulting in a more robust biomarker candidate than the traditional 1/f slope, and provides a computational methodology for generating biologically plausible neurophysiological signals.

\end{abstract}

\begin{figure}[htbp]
  \centering
  \includegraphics[width=\linewidth]{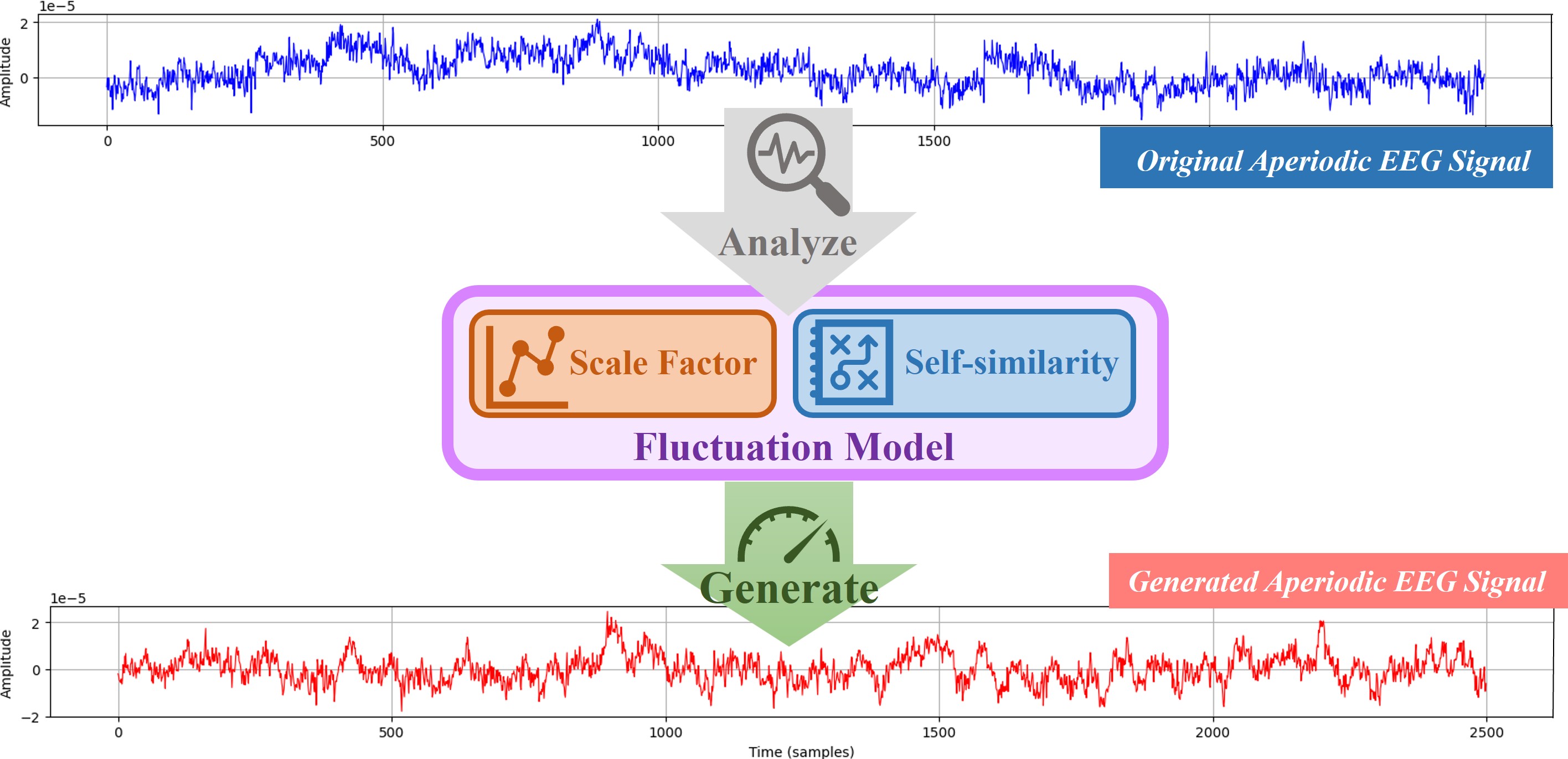}
  \caption{\textbf{Fluctuation-based generation of aperiodic EEG signals.} The original aperiodic EEG signal (top) is analyzed to extract only 2 parameters, scale factor and self-similarity, to guide the fluctuation model to generate a synthetic signal (bottom) with similar scale-free dynamics.}
  \label{fig:abstract}
\end{figure}

\section{Introduction}

Electrophysiological brain signals, such as Electroencephalography (EEG), comprise both periodic and aperiodic components~\citep{berger1929elektroenkephalogramm}.
These components can be separated and analyzed in the frequency domain using power spectral density (PSD) methods~\citep{FOOOF}. 
While oscillatory activity has been widely investigated, the aperiodic component—traditionally regarded as background noise—has recently garnered attention for its potential to convey meaningful information.
Emerging evidence suggests that the aperiodic component reflects a range of neurocognitive and physiological states, including cognitive processing speed~\citep{1onf_cognitive}, neurological disorders~\citep{1onf_disorder}, and levels of consciousness~\citep{1onf_consciousness}.

The aperiodic spectral trend is commonly modeled in the spectral domain by a $1/f^{\beta}$ power law, wherein spectral power decays with frequency according to an exponent $\beta$~\citep{scale-free-review}.
Although this power-law behavior is empirically well-established, the neural and computational origin remains a subject of ongoing debate.
One prominent theoretical hypothesis is the neural avalanche theory, which attributes the $1/f$ scaling law in spectral domain to scale-invariant cascades of synchronized neuronal activity over short timescales, posited to arise from self-organized criticality in neural systems~\citep{so-critic,neuro_avalanche}.
An alternative framework, known as dynamical criticality, is also proposed to account for the foundation of aperiodic power spectral~\citep{chaudhuri2018random}. It proposes that this behavior emerges near a bifurcation point in neural dynamics, where the system exhibits critical slowing down and increased sensitivity to perturbations.
Another hypothesis, the synaptic scaling hypothesis, also supports the source of $1/f$ is its dynamics, which suggests that the $1/f$ power-law behavior arises from a temporal hierarchy of synaptic responses~\citep{miller2009power}. 
Notably, such an explanation does not necessarily require the presence of a critical state, where the existence of a critical state in neural systems is also a subject of ongoing debate~\citep{not_critical_hypothesis}. 

Despite the diversity of these theories, one crucial and often overlooked feature of aperiodic activity is its inherently stochastic nature.
The fluctuations in aperiodic EEG signals can be interpreted as spontaneous neural sampling process~\citep{fns}.
In many modeling approaches, these fluctuations are represented as white noise within stochastic differential equations~\citep{wilson_cowan}.
While theoretically tractable and self-consistent, this simplified assumption may fail to capture aperiodic components observed in real brain signals.
This raises a question: Can stochastic fluctuations in brain signals be parameterized using a statistical model, and if so, how do these parameters relate to the properties of aperiodic activity?

In this study, we propose a framework that conceptualizes the aperiodic component of EEG signals as emerging from intrinsic neural fluctuations. Specifically, we introduce two key parameters—self-similarity and scale factor—to quantitatively characterize these neural fluctuations.
Our results demonstrate that this parameterization enables the reconstruction of EEG-like signals that preserve core dynamical properties, including the 1/f spectral profile and long-range dependency. Additionally, we showed that, the self-similarity exponent in neural fluctuations is crucial for the generation of 1/f signal and is related to long range dependency. Furthermore, we show that the distribution of neural fluctuation varies over time, and within parts of a time window, neural fluctuations exhibit non-Gaussian behavior, suggesting a more complex structure than traditionally assumed. Our findings offer new insights into the generative mechanisms of aperiodic EEG activity and contribute to a more comprehensive understanding of brain dynamics beyond oscillatory models.

\section{Related Works}

\subsection{Theories for aperiodic 1/f spectral}

The presence of 1/f spectral characteristics in EEG signals has been extensively documented~\citep{scale-free-review}.
In addition to spectral analysis, detrended fluctuation analysis (DFA) has been introduced as a complementary method to quantify the long-range temporal correlations indicative of 1/f behavior in neural data~\citep{DFA}.
While spectral analysis focuses on frequency-domain properties, DFA provides insights into long-range dependencies in the time domain.
Several theoretical frameworks have been proposed to explain the emergence of 1/f dynamics, many of which converge on the concept of the critical state in neural systems~\citep{critical_review}.
A critical state refers to the condition in which the system operates near a phase transition point, and is linked to the emergence of complex dynamics and long-range temporal correlations~\citep{so-critic,critical_review}.
There are multiple approaches to modeling or achieving a critical state in neural systems. Self-organized criticality relies on local interactions and activity-dependent adaptation mechanisms that inherently drive the system toward a critical point~\citep{anderson_crit}. Alternatively, dynamical criticality models focus on tuning system parameters near bifurcation points where small perturbations can lead to large-scale, system-wide changes. Both approaches have been shown to give rise to activity patterns that resemble 1/f dynamics in their statistical properties~\citep{chaudhuri2018random}.
In this paper, we focus not on the mechanisms that generate criticality per se, but rather on its observable consequences—specifically, the statistical characteristics of neural fluctuations derived from electrophysiological signals. We investigate whether these fluctuations can serve as indicators of aperiodic neural activity and reflect key parameters governing the underlying dynamical regime.

\subsection{Neural Fluctuations}

Modeling large-scale neural dynamics frequently involves incorporating stochastic fluctuations to account for variability around mean firing rates or field potentials. A prevalent assumption in such models is that these fluctuations are temporally uncorrelated, often represented by white noise or Poisson processes in spiking models. This assumption is rooted in mathematical self-consistency: fluctuations over a given time interval are treated as the aggregate of fluctuations over smaller, independent sub-intervals. Consequently, temporally independent Gaussian noise—i.e., white noise—is often adopted as the default model for modeling neural variability~\citep{wilson_cowan}.
However, this assumption has been challenged by alternative theoretical frameworks. Fractal neural sampling theory~\citep{fns} argues that neural stochasticity should be modeled as a self-similar process, governed by long-tailed Lévy distributions rather than Gaussian ones. This theory aligns with empirical evidence of non-Gaussian, long-range dependencies in neural activity.
Further theoretical support comes from studies in statistical noise modeling, particularly those based on fractional Brownian motion (fBm), which provide a principled framework for modeling fluctuations that exhibit both self-similarity and long-range temporal correlations~\citep{fBm}.
In this paper, we further investigate the neural fluctuation models that capture the distributions extracted from brain signals, and offer a different view of neural dynamics.

\section{Methods}

\begin{figure}[htbp]
  \centering
  \includegraphics[width=\linewidth]{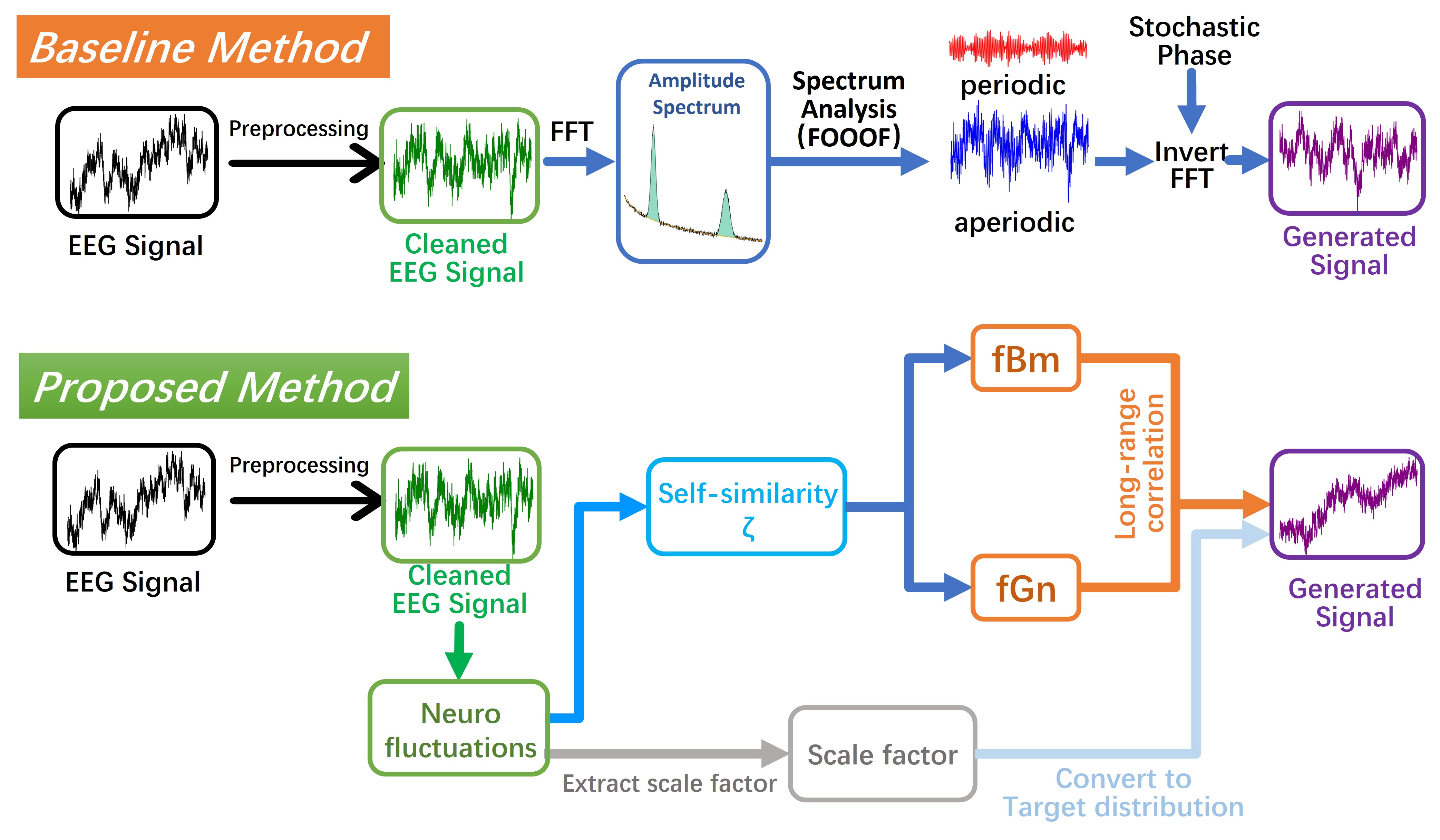}
  \caption{\textbf{The architecture of the fluctuation analysis.} Fluctuation analysis involves the extraction of fluctuation patterns from a given signal, followed by the computation of two fundamental parameters: the scale factor $s$ and the self-similarity exponent $\zeta$. The exponent $\zeta$ is subsequently transformed and used in synthesis signals exhibiting long-range correlations, generated through fractal Brownian motion (fBm) and fractal Gaussian noise (fGn). Finally, the distribution of the generated signal is reshaped according to the fluctuation distribution and scale factor $s$.}
  \label{fig:architecture}
\end{figure}

\subsection{Aperiodic Components Extraction}
To separate the periodic and aperiodic components of the EEG signal~\citep{baseline}, we apply a combination of Fourier transform and the FOOOF (Fitting Oscillations and One Over F) model~\citep{FOOOF}. First, we perform the Fast Fourier Transform (FFT) to convert the EEG signal into the frequency domain, yielding both the amplitude and phase spectra.
\begin{equation}
X(k) = \sum_{i=0}^{N-1} x(i) \, e^{-i 2\pi \frac{kn}{N}}, \quad k = 0, 1, \ldots, N-1
\end{equation}
where $x(i)$ represents the time-domain signal, and N is the total number of samples. From the FFT, we compute the Power Spectral Density (PSD) as:
\begin{equation}
PSD(f) = \frac{1}{N} \left| \text{X}(f) \right|^2
\end{equation}
which is then filtered to retain with frequencies $f>1$Hz. Next, we apply the FOOOF model to PSD, which parameterizes the aperiodic background and identifies superimposed oscillatory peaks. The periodic component, consisting of Gaussian-shaped spectral peaks, reflects narrowband oscillatory activity. We extract the parameters of each peak (center frequency and bandwidth), and for peaks with center frequencies below 50 Hz, we replace their spectral contributions with the corresponding values from the aperiodic fit. This process allows us to reconstruct a purely aperiodic spectrum while preserving the original phase information. Finally, the modified frequency spectrum which now reflects only the aperiodic component is combined with the phase spectrum to reconstruct the aperiodic signal in the time domain:
\begin{equation}
x_{\text{aperiodic}}(t) = \mathcal{F}^{-1} \left( \left| \hat{X}(f) \right| \cdot e^{i \phi(f)} \right)
\end{equation}
where $\hat{X}(f)$ is modified FFT magnitude, $\phi(f)$ is phase spectrum, and $\mathcal{F}^{-1}$ denotes the inverse FFT.

\subsection{Neural Fluctuations}

\textbf{Definition.} We regard fluctuations as the stochastic variations in a signal~\citep{fns}. Given an aperiodic signal $x(t)$, we define the fluctuation $\Delta_{\tau}(t)$ over a time lag $\tau$ as the random variable:

\begin{equation}
\Delta_{\tau}(t) = P(x(t+\tau) - x(t))
\end{equation}

and we model its distribution as $\Delta_{\tau}\sim P_\tau$, where $P_\tau$ denotes the probability distribution of the temporal difference at time lag $\tau$.

\textbf{Extraction.} Since fluctuations are defined with respect to the aperiodic component of the signal, we treat the observed signal as a realization drawn from the underlying fluctuation distribution $P_\tau$. Specifically, we compute incremental differences across multiple time lags $\tau$, selected on a logarithmic scale. These temporal increments are regarded as empirical samples of the fluctuation variable $\Delta_{\tau}(t)$, and thus serve as the basis for estimating the statistical properties of $P_\tau$.

\textbf{Distributional Analysis.} To characterize the underlying statistical properties of the fluctuations, we investigate which probability distribution best fits the empirical increments. Several candidate distributions are considered, including the normal, logistic, and Cauchy distributions. The goodness of fit is evaluated using the Kolmogorov–Smirnov (K–S) test~\citep{ks}, with a p-value greater than 0.05 indicating an acceptable fit at the $5\%$ significance level.

\textbf{Scale factor.} The scale factor, denoted by $s$, is defined as the range of the distribution $P_\tau$. In the normal and logistic distributions, $s$ corresponds to the standard deviation and the scale parameter, respectively. In practice, we estimate the scale factor using maximum likelihood estimation (MLE) method ~\citep{scipy}. 

\textbf{Self-similarity.} We investigate the power-law scaling ~\citep{lrtc} behavior of distribution parameters across varying time lags $\tau$. 
In self-similar processes, the scale of fluctuations increases with the time lag according to a power-law relationship.
Accordingly, scale factor $s$ and self-similarity exponent $\zeta$ are modeled and fitted using a power law function:
\begin{equation}
s(\tau) = C \cdot \tau^\zeta
\end{equation}
where C represents a constant. In the case of temporally uncorrelated fluctuations, i.e, white noise, the expected self-similarity exponent is $\zeta = 0.5$, corresponding to a square-root growth of the fluctuation scale with respect to time lag.

\subsection{Stochastic Generation}
Neural fluctuations are generated using only two parameters: the scale factor $s$ along with their corresponding self-similarity exponent $\zeta$. 
Based on self-similarity exponent $\zeta$, we derive the corresponding Hurst exponent in fractional Gaussian noise (fGn) and fractional Brownian motion (fBm)~\citep{fBm}.
Using this relationship, we simulate long-range temporally correlated neural fluctuations employing either fGn or fBm. 

\begin{equation}
Hurst = 
\begin{cases}
\alpha, & \text{if } \alpha < 1 \quad (\text{fGn})\\
\alpha - 1, & \text{if } \alpha \geq 1 \quad (\text{fBm})
\end{cases}
\end{equation}

where $\alpha = 2 \cdot \zeta + 0.5$.
Fractional noise samples are generated using the Cholesky method as implemented in the fbm library, for computational stability~\citep{cholesky,fbm_lib}.

To ensure consistency with the marginal distribution of empirical EEG signals, the generated time series are first standardized to have zero mean and unit variance. Then, we transform these standardized values into a uniform distribution via the cumulative distribution function (CDF) of the standard normal distribution. Subsequently, to obtain a time series exhibiting both the desired distributional properties and long-range temporal dependencies, the uniform values are mapped to the target marginal distribution using the inverse CDF (percent-point function) of either the logistic or normal distribution, as determined by the chosen generative model. The inverse CDF is obtained solely based on the scale factor $s$, given that the fluctuation distribution has a mean of zero, both empirically and theoretically. Finally, the transformed sequence is cumulatively summed to yield the resulting stochastic neural fluctuations.

\section{Experiments}
\subsection{Dataset and Pre-processing}
\textbf{Dataset.} We utilized the publicly available eyes-open resting-state EEG dataset, which included EEG recordings and cognitive data from 71 participants under normal sleep (NS) and sleep deprivation (SD) conditions~\citep{EEG_dataset}. EEG signals were recorded at a sampling rate of 500 Hz for 300 seconds per session. To minimize potential order effects, the order of sessions was counterbalanced across participants. In this study, we primarily focused on investigating the characteristics of the 1/f component in spontaneous neural activity using eyes-open resting-state EEG data from the normal sleep session. After excluding two participants with incomplete eyes-open recordings, data from 69 participants were retained for analysis. 

\textbf{Pre-processing.} We implemented a standardized and automated EEG preprocessing pipeline~\citep{FieldTrip} based on MNE-Python~\citep{MNE}. Eyes-open recordings from 69 participants were included in the analysis. Signals were high-pass filtered at 1 Hz to remove slow drifts, re-referenced to the common average, and notch filtered at 50 Hz to suppress power line interference. Independent Component Analysis (ICA) was conducted using the Infomax algorithm~\citep{Infomax} with 15 components. Artifactual components, including those reflecting ocular and muscular activity, were automatically identified via the ICLabel classifier~\citep{MNE-ICALabel, ICALabel}, and only components labeled as "brain" or "other" were retained for subsequent analysis. We focused on the occipital channels O1, O2, and Oz, given their established involvement in visual processing and vigilance. The continuous EEG data from each channel were segmented into 60 non-overlapping epochs of 5 seconds, yielding temporally consistent inputs for downstream analysis.

\subsection{Baseline}
As a baseline, we employ a widely adopted Fourier-based method to synthesize aperiodic signals with power-law spectral properties slope and offset (Slope-FPG)~\citep{baseline, baseline—power-law}. Specifically, we first transform the EEG signals into the frequency domain using FFT to extract both the amplitude and phase spectra. The aperiodic component of the amplitude spectrum is then parameterized using FOOOF algorithm~\citep{FOOOF}, which estimates the slope and offset of the aperiodic background activity. Using these estimated parameters, we reconstruct a synthetic amplitude spectrum that preserves only the aperiodic component. Combining the reconstructed amplitude spectrum with a random phase spectrum, we apply the inverse FFT to generate time-domain signals that reflect the aperiodic structure of the EEG while discarding oscillatory components.

\subsection{Evaluation Metrics}
To comprehensively evaluate the fidelity of reconstructed aperiodic signals, we employ four quantitative metrics to compare the original and reconstructed signals, each capturing distinct signal properties:
\textbf{Slope.} Captures the slope $\beta$ of the aperiodic background component in the Power Spectral Density (PSD);
\textbf{Alpha.} Quantifies temporal self-similarity through Detrended Fluctuation Analysis~\citep{DFA};
\textbf{Spectra Cosine Similarity.} Evaluates the overall structural resemblance of the original and reconstructed spectra;
\textbf{Normalized Mutual Information (NMI).} Quantifies the mutual information encompassing both linear and nonlinear dependencies between the original and reconstructed spectra.

We deliberately exclude error-based metrics such as Root Mean Square Error (RMSE), as our goal is not to reproduce the original signal point-by-point, but rather to preserve its key temporal and spectral properties.
Collectively, these metrics encompass two analytical dimensions: long-range dependency and spectral similarity.

\section{Results}

\subsection{Fluctuation modeling}

\begin{wrapfigure}[10]{R}{0.5\textwidth}
  \vspace{-80pt}
  \centering
  \includegraphics[width=0.9\linewidth]{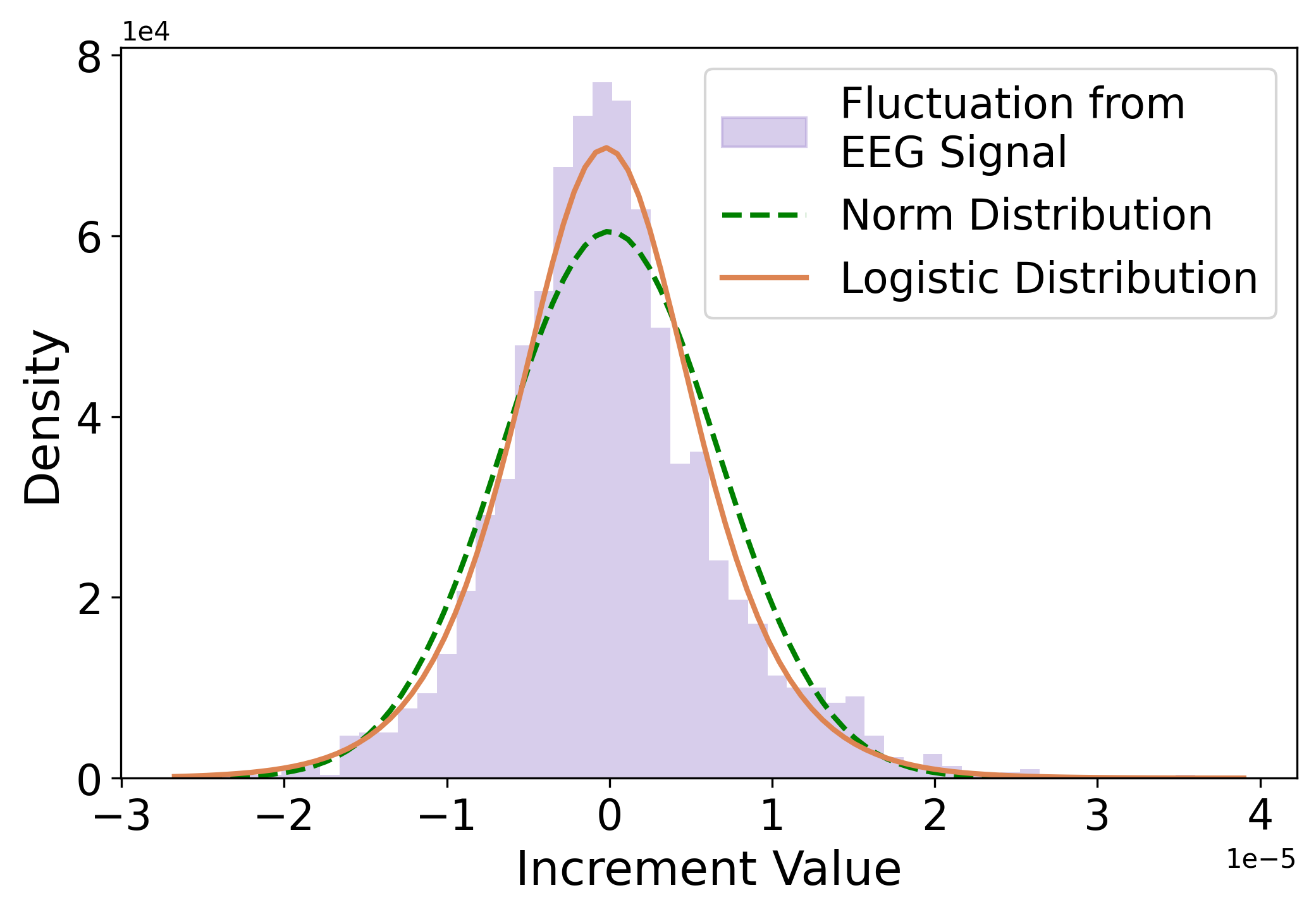}
  \vspace{-10pt}
  \caption{\textbf{Distribution of fluctuations $P_{\tau}$ modeled using the Logistic and Normal distributions.} While both distributions exhibit a bell-shaped curve, the Logistic distribution provides a better fit for the data.}
  \label{fig:distribution}
\end{wrapfigure}

First, we performed data pre-processing and extracted the neural fluctuations of interest. Each segment of the distribution $P_{\tau}$ was then analyzed to examine and validate the commonly used hypothesis that the fluctuations follow a Gaussian distribution. The results of the statistical tests are presented in Table~\ref{tab:distribution_fit}. Time lag $\tau$ is measured in ticks; given the dataset's sampling frequency of 500 Hz, each tick corresponds to a $1/500$ second.

The findings indicate that a significant portion of the extracted fluctuations deviate from a Gaussian distribution, contrasting sharply with theoretical predictions. In addition to the Gaussian model, we evaluated the fit of the Lévy distribution—a type of stable distribution characterized by heavy long tails—commonly used to model extreme events resulting from rare, large jumps in the signal. The result showed that $P_{\tau}$ is not a Levy distribution.

We also examined alternative distributions and found that the logistic distribution provided the best fit. Although both the logistic and normal distributions are symmetric and bell-shaped, the logistic distribution exhibits heavier tails. Notably, the logistic distribution is not stable, implying that the sum of independent, identically distributed logistic variables does not retain the same distributional form. However, our results reveal that the sum of logistic-distributed fluctuations over short time lags tends to approximate a logistic distribution over longer time lags. This observation suggests the presence of underlying neural mechanisms, which warrant further investigation.

\begin{table}[htbp]
    \centering
    \caption{\textbf{Goodness-of-fit (\%) of different distributions across differencing Time Lag $\tau$}}
    \begin{tabular}{cccccccc}
    \toprule
    $\tau$ & Norm & Lévy & Student-t & Gamma & Laplace  & Cauchy  & Logistic \\
    \midrule
    1  & 52.69 & 0.00 & 6.47  & 41.06 & 4.12 & 0.16 & \textbf{87.71} \\
    2  & 59.81 & 0.00 & 4.60  & 46.59 & 2.78 & 0.11 & \textbf{90.18} \\
    4  & 87.61 & 0.00 & 3.69  & 78.66 & 1.01 & 0.06 & \textbf{91.40} \\
    8  & 89.23 & 0.00 & 3.35  & 83.79 & 0.91 & 0.02 & \textbf{89.25} \\
    16 & 83.95 & 0.00 & 4.79  & 80.16 & 1.41 & 0.02 & \textbf{89.34} \\
    32 & 76.28 & 0.00 & 8.02  & 70.76 & 2.46 & 0.02 & \textbf{87.50} \\
    64 & 67.31 & 0.00 & 12.71 & 62.88 & 3.91 & 0.06 & \textbf{84.49} \\
    \bottomrule
    \end{tabular}
    \label{tab:distribution_fit}
\end{table}

\begin{figure}[htbp]
  \centering
  \includegraphics[width=0.9\linewidth]{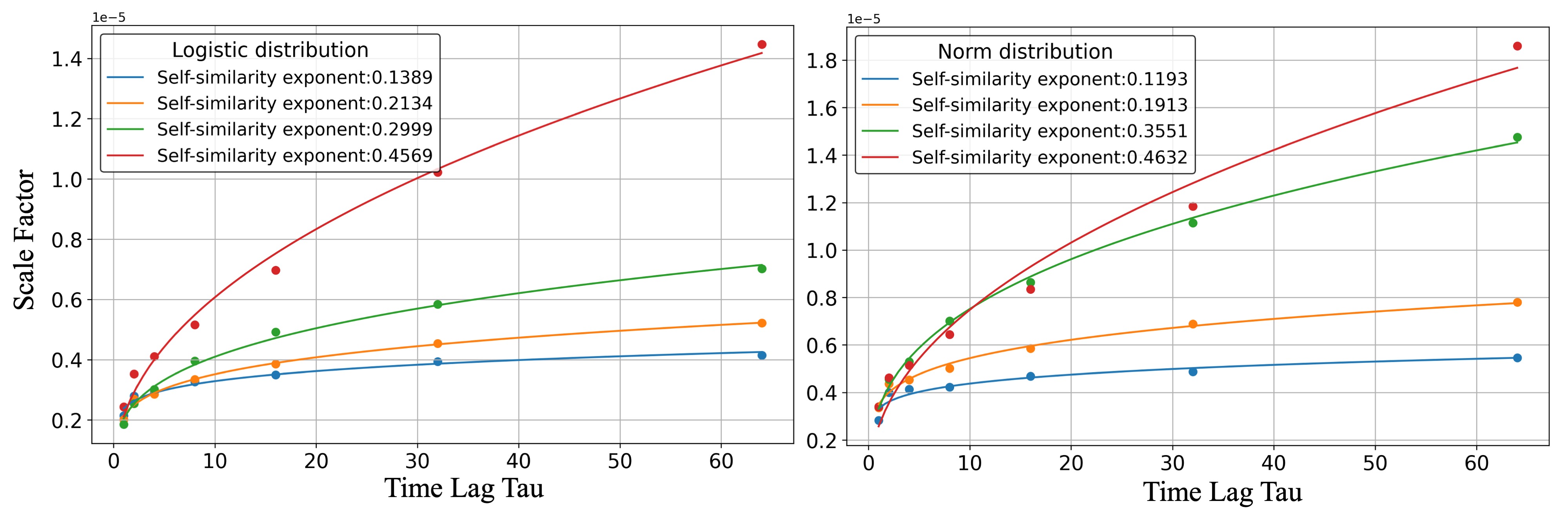}
  \caption{\textbf{The increase in fluctuation scale follows a power-law relationship.} For each segment, the fluctuation scale $s$ is computed across various time lags $\tau$. The figure illustrates representative examples of the power-law fitting of the self-similar exponent $\zeta$, using the Logistic distribution (left) and the Normal distribution (right). The scaling behavior of the fluctuations exhibits a power-law dependence, indicating an underlying self-similarity. Despite using different distribution, the increase of scale factor $s$ have a similar trend. We choose segments that demonstrating varying degrees of self-similar behavior with different $\zeta$. }
  \label{fig:power_law}
\end{figure}

Next, we investigate the self-similarity of fluctuations by estimating the power-law exponent characterizing the scaling behavior of their distribution. Specifically, we analyze the self-similarity within each signal segment, fitting both normal and logistic distributions to the data.
As illustrated in Fig.~\ref{fig:power_law}, our results demonstrate that the magnitude of fluctuations increases according to a power law, indicative of self-similarity behavior. Notably, the fitted exponents reveal considerable variability in self-similarity across different signal segments. In some segments, the self-similarity exponent is close to $\zeta = 0$, indicating an absence of scaling behavior, whereas in others, $\zeta$ approaches 0.5, corresponding to nearly uncorrelated noise.

\begin{figure}[htbp]
  \centering
  \includegraphics[width=0.8\linewidth]{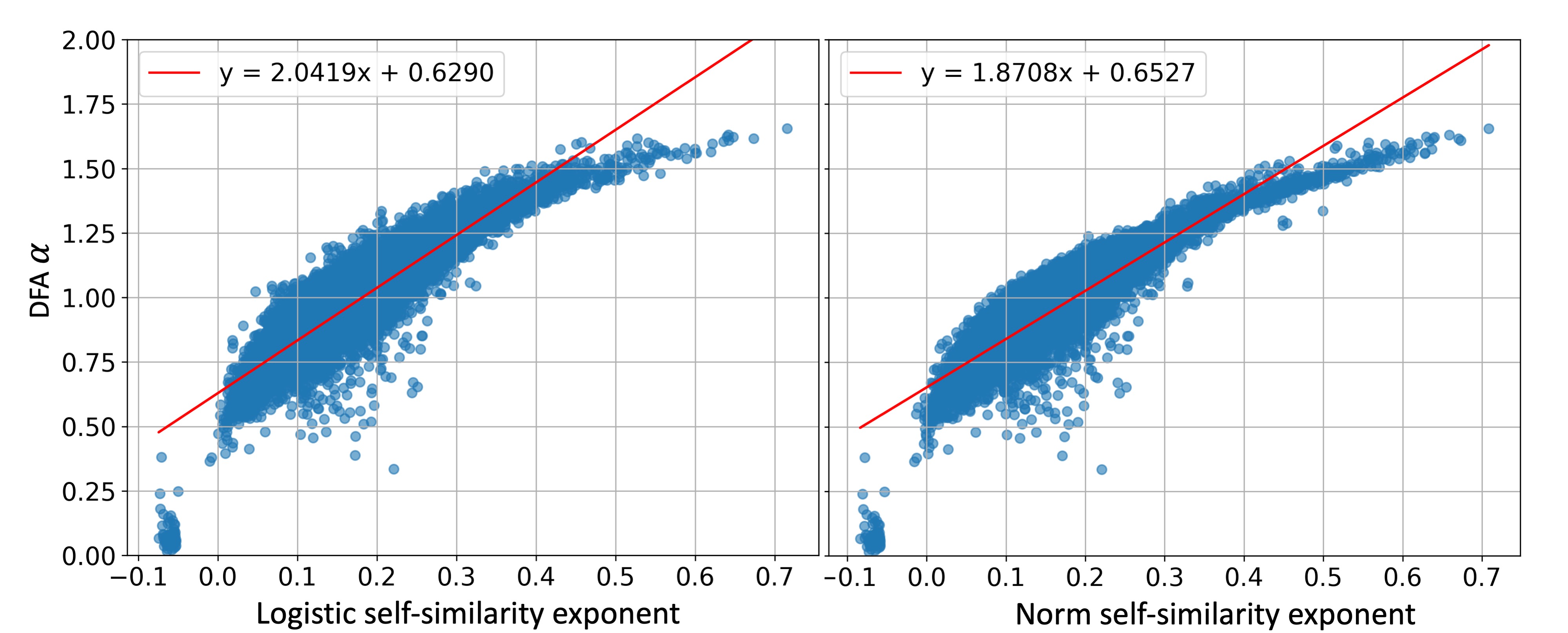}
  \caption{\textbf{Relationship between the self-similarity exponent and long-range dependency.} The self-similarity exponent $\zeta$ is computed using both the normal (Left) and logistic distributions (Right), while long-range dependency is assessed via DFA. The observed correlation between $\zeta$ and the DFA scaling exponent $\alpha$ suggests that $\zeta$ can serve as an indicator of long-range dependency in EEG signals.}
  \label{fig:lr_dependency}
\end{figure}

To further explore the presence of long-range dependencies associated with self-similarity, we applied Detrended Fluctuation Analysis (DFA), a well-established method for quantifying long-range correlations in time series data~\citep{DFA}. As shown in Fig.~\ref{fig:lr_dependency}, we observe a strong correlation between the self-similarity exponent $\zeta$ and the DFA scaling exponent $\alpha$. This relationship was further confirmed through linear regression analysis.
Although minor differences in linear fit were observed depending on whether the normal or logistic distribution was used to model the fluctuations, the overall trend remains robust. These findings suggest that the self-similarity exponent $\zeta$ serves as a reliable indicator of long-range dependency in EEG signals.

\subsection{EEG signal generation}

We employ a comprehensive set of metrics to evaluate the quality of signal generation. As a baseline, we use a Fourier-based method that synthesizes aperiodic EEG signals directly from the power spectrum, incorporating explicit information about the 1/f spectral slope. This approach is commonly adopted for aperiodic EEG signal generation~\citep{baseline}.

\begin{figure}[htbp]
  \centering
  \includegraphics[width=\linewidth]{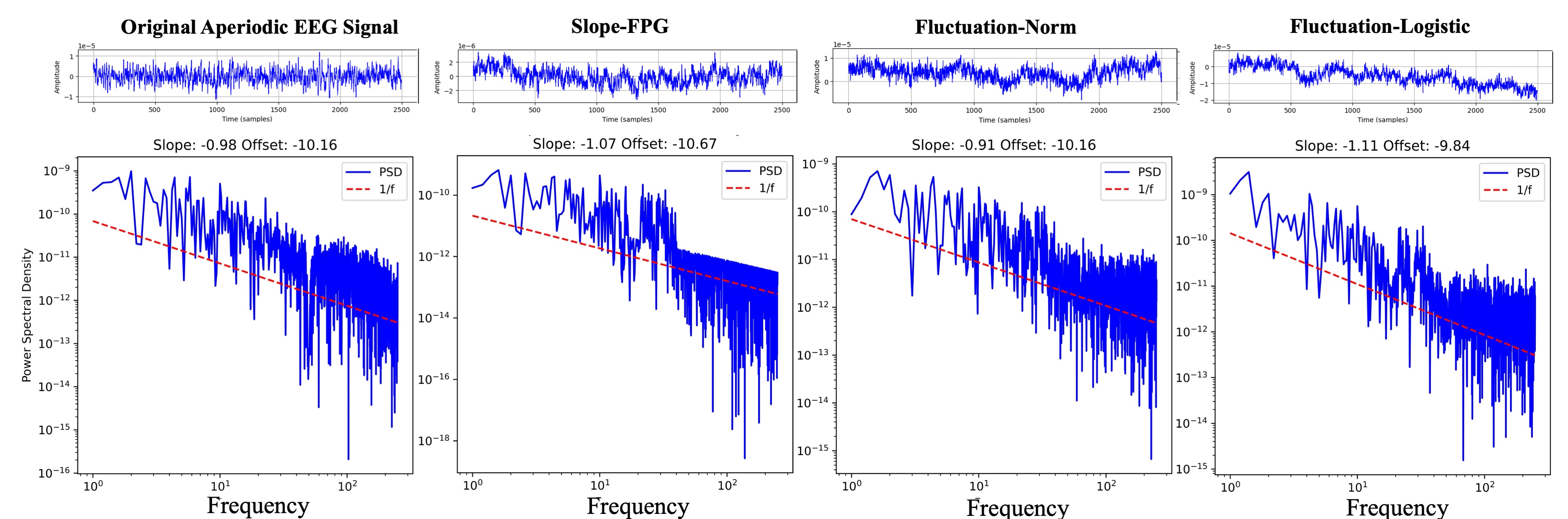}
  \caption{\textbf{An example of aperiodic EEG signal generation using different methods.} Upper: Aperiodic EEG signal. Lower: PSD and 1/f fitting using FOOOF~\citep{FOOOF}. All methods produce valid EEG-like fluctuations with visible 1/f characteristics. While the slope derived from the slope-FGN method is closer to that of the original data, our proposed method yields a PSD shape that more closely resembles the overall spectral profile of real EEG signals.}
  \label{fig:gene_result}
\end{figure}

Our results demonstrate that the signals generated by our model exhibit power spectra closely resembling those of the original signals, as evidenced by a slightly higher cosine similarity.
While the logistic distribution provides a better fit to the empirical fluctuation distributions, the normal distribution can also be employed for fluctuation modeling, albeit with a marginally lower spectral similarity. This finding suggests that self-similarity plays a critical role in determining spectral similarity.
Notably, our fluctuation-based approach does not involve any Fourier transform operations and, therefore, lacks direct access to the slope information. Consequently, the generated spectral slope deviates more from the original signal compared to the baseline method. This is expected given that the baseline explicitly reconstructs the signal using the known spectral slope.
Additionally, our method shows a clear advantage in capturing long-range temporal dependencies. This advantage can be attributed to the use of self-similarity mechanisms within our model, which are inherently related to such dependencies.

\begin{table}[htbp]
\centering
\caption{\textbf{Comparison of signal characteristics with different generation methods}}
\begin{tabular}{c|c|ccc}
\toprule
\textbf{Model} & \textbf{DFA $\alpha$ Diff. (\%)$\downarrow$} & \textbf{Slope Diff.(\%)$\downarrow$}  & \textbf{Cosine Sim. $\uparrow$} & \textbf{NMI $\uparrow$} \\
\midrule
\textit{} & \textit{Long-range Dependency} & \multicolumn{3}{c}{\textit{Spectral Similarity}} \\
\midrule
Slope-FPG    & 31.10  & \textbf{19.10}  & 0.71 & \textbf{0.56} \\
Fluc.-Norm (\textbf{Ours})      & 17.59  & 44.87  & 0.72 & 0.54 \\
Fluc.-Logistic (\textbf{Ours})  & \textbf{17.11}  & 33.97  & \textbf{0.75} & 0.51 \\
\bottomrule
\end{tabular}
\label{tab:signal_characteristics}
\end{table}

Given that EEG is a highly noisy signal, we also evaluate the quality of the spectral fitting of the generated signals. Despite the baseline achieving a more accurate slope fit, it exhibits a higher overall fitting error in the power spectrum. This suggests that our model provides a more faithful reconstruction of the entire spectral profile, rather than merely fitting a predefined linear slope. 

In summary, fluctuation modeling methods can be used to generate an aperiodic EEG signal, which captures the long range dependency and spectral property of the original signal. However, the slope of 1/f may not be reproduced exactly, since fitting of a slope relies highly on both the aperiodic fitting algorithm and the smoothness of the PSD.

\begin{table}[htbp]
\centering
\caption{\textbf{Comparison of slope fitting quality across generation methods}}
\begin{tabular}{lcc}
\toprule
\textbf{Model} & \textbf{Slope $R^2$ $\uparrow$} & \textbf{Overall Error $\downarrow$} \\
\midrule
Slope-FPG         & \textbf{0.45} & 0.58 \\
Fluc.-Norm (\textbf{Ours})        & 0.28 & \textbf{0.45} \\
Fluc.-Logistic (\textbf{Ours})    & 0.31 & \textbf{0.45} \\
\midrule
original EEG signal & 0.55 & 0.44 \\
\bottomrule
\end{tabular}
\label{tab:slope_metrics}
\end{table}

\section{Conclusion}

In conclusion, this study introduces two key parameters—self-similarity and scale factor—as quantitative tools for characterizing neural fluctuations. Through this parameterization, we successfully reconstructed EEG-like signals that preserved essential spectral features, notably the characteristic 1/f profile. Our findings highlight the critical role of the self-similarity exponent in generating 1/f dynamics, emphasizing its importance in understanding aperiodic brain activity. Additionally, we observed temporal variations in the statistical properties of neural fluctuations, with aperiodic component can be better described using Logistic distribution rather than Normal distribution. Together, these insights advance our understanding of the generative mechanisms behind aperiodic EEG activity and offer a broader perspective on brain dynamics.

\section{Discussion and Limitation}

Our study presents a novel and promising approach for identifying a potential biomarker based on electroencephalography (EEG) signals. Although EEG data—particularly the power spectral density (PSD) and its slope—are inherently noisy and often require extended recording durations for effective denoising, our findings indicate that reliable parameter estimation can be achieved using fluctuation modeling. 
In this work, we only use signal segments as brief as approximately five seconds separately, without averaging through segments.
Despite the short duration of these recordings, the parameter fitting exhibited high consistency, suggesting that the proposed method may serve as a robust biomarker, especially for neurological conditions or cognitive states associated with disorders in PSD slope.

However, it is important to acknowledge that the present analysis was conducted exclusively on an eyes-open EEG dataset. Consequently, the potential effects of cognitive state on the observed neural dynamics were not examined. Future research should incorporate a broader range of cognitive conditions to evaluate the generalization and stability of the proposed biomarker across different mental states.
Furthermore, the applicability of this method may extend beyond EEG. It would be valuable to investigate its effectiveness in other modalities of brain signals, particularly those characterized by high temporal resolution and long-range temporal dependencies, similar to EEG.

Lastly, our results offer new insights into the underlying sources of neural activity. The observed characteristics of the parameter distributions—such as logistic-like behavior and self-similarity—suggest the presence of structured and possibly hierarchical neural mechanisms. These findings may contribute to the development of biologically plausible models of neural dynamics and improve our understanding of the fundamental principles governing brain function.

\newpage

\bibliography{main}
\bibliographystyle{abbrvnat}


\newpage
\appendix
\startappendix

\section{Appendix}

\subsection{Mathematical explanation of fluctuation modeling}

This appendix presents a detailed mathematical exposition of the relationship between the self-similarity exponent and the Hurst exponent in the context of fluctuation modeling. It further elucidates how this relationship informs various sampling methodologies for generating synthetic signals that exhibit specific self-similar characteristics.





\subsubsection{Self-Similarity Exponent in Fluctuation Analysis}

In our fluctuation analysis framework, we characterize self-similarity through the exponent $\zeta$, which describes how the scale of fluctuations grows with increasing time lag $\tau$:

\begin{equation}
s(\tau) = C \cdot \tau^\zeta
\end{equation}

\noindent where $s(\tau)$ is the scale parameter of the fluctuation distribution at time lag $\tau$, and $C$ is a constant. This power-law relationship is a direct manifestation of self-similarity.

The relationship between our self-similarity exponent $\zeta$ of fluctuations and the Generalized Hurst exponent $\alpha$ of a stochastic process is:

\begin{equation}
\alpha = 2 \zeta + 0.5
\end{equation}

This formulation enables the characterization of a wide range of stochastic processes:
For temporally uncorrelated fluctuations where $\zeta = 0.5$, it corresponds to $\alpha = 1.5$ where the process is a Brownian motion process. Additionally, for non-incrementing fluctuation where $\zeta = 0$, resulting in a process is a white noise where $\alpha = 0.5$.

This relationship is critical for understanding how our measured self-similarity parameter $\zeta$ can be used to generate synthetic signals with appropriate long-range dependency properties.

\subsubsection{Generating Self-Similar Processes}

To generate signals with specific self-similarity properties, we utilize the relationship between $\zeta$ and $\alpha$ in conjunction with different sampling methods. Our approach focuses on two mathematical constructs: fractional Gaussian noise (fGn) and fractional Brownian motion (fBm).
Since the practical generation of such processes relies on the Hurst exponent $H$, instead of generalized Hurst exponent $\alpha$, we map $\alpha$ to Hurst exponent $H$ on the appropriate model (fGn or fBm) based on the value of $\alpha$:

\begin{equation}
\text{Hurst exponent for generation} \quad H = 
\begin{cases}
\alpha, & \text{if } \alpha < 1 \quad (\text{use fGn})\\
\alpha - 1, & \text{if } \alpha \geq 1 \quad (\text{use fBm})
\end{cases}
\end{equation}

This mapping ensures the appropriate choice of model based on the desired degree of self-similarity and long-range correlation structure.

\subsubsection{Non-Gaussian Properties in Self-Similar Sampling}

An important aspect of our work is the finding that neural fluctuations often follow non-Gaussian distributions, specifically the logistic distribution:

\begin{equation}
f(x; \mu, s) = \frac{e^{-(x-\mu)/s}}{s(1+e^{-(x-\mu)/s})^2}
\end{equation}

This distribution has heavier tails than the Gaussian distribution, better capturing the occasional large deviations observed in neural signals.

When generating synthetic signals, we must carefully preserve both the long-range dependency structure characterized by the Hurst exponent (derived from $\zeta$) and the non-Gaussian distributional properties characterized by the scale factor $s$.
This dual preservation is accomplished through our two-step process:
1) Generate a Gaussian process with the required correlation structure using the Cholesky method.
2) Transform the marginal distribution while preserving the dependence structure.
This approach allows us to create synthetic signals that faithfully reproduce both the temporal dependency structure and the distributional properties of real neural fluctuations.

\subsubsection{From Fluctuations to Time Series}

After generating fluctuations with the appropriate self-similarity and distributional properties, we obtain the final time series through cumulative summation:

\begin{equation}
X(t) = \sum_{i=1}^{t} \Delta(i)
\end{equation}

\noindent where $\Delta(i)$ represents the generated fluctuations.

This process results in a synthetic time series that exhibits the characteristic 1/f-like behavior in its power spectrum and the long-range temporal dependencies that are observed in real EEG signals.

In summary, the relationship between the self-similarity exponent $\zeta$ and the Hurst exponent $H$ provides a critical link between our empirical fluctuation analysis and the generation of synthetic signals with appropriate fractal properties. By carefully translating between these parameters and incorporating the non-Gaussian nature of neural fluctuations, our approach offers a mathematically principled method for modeling the complex stochastic processes underlying aperiodic brain activity.

\newpage

\subsection{Additional Experimental Results}

\begin{figure}[htbp]
  \centering
  \includegraphics[width=\linewidth]{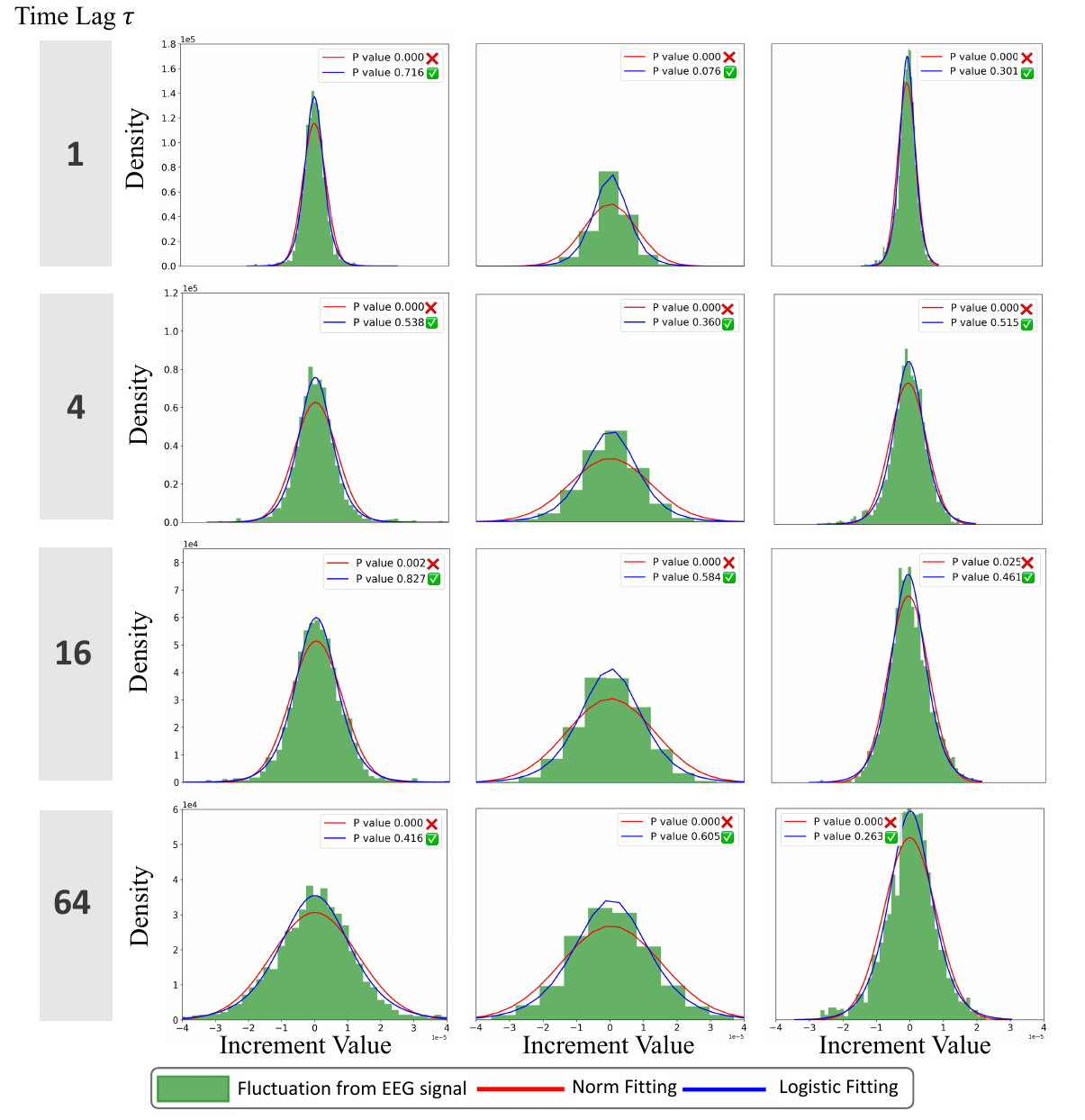}
  \caption{\textbf{Distribution of fluctuations $P_\tau$ fitting to Normal and Logistic distributions across different time lags.} The Kolmogorov–Smirnov (KS) test is employed to evaluate the goodness of fit between the distribution of fluctuations $P_\tau$ and theoretical models, as illustrated in the figure. A higher p-value obtained from the KS test indicates that the null hypothesis cannot be rejected, suggesting no statistically significant difference between the theoretical distribution and the empirical data.}
  \label{fig:appendix_distribution}
\end{figure}

\begin{figure}[t]
  \centering
  \includegraphics[width=\linewidth]{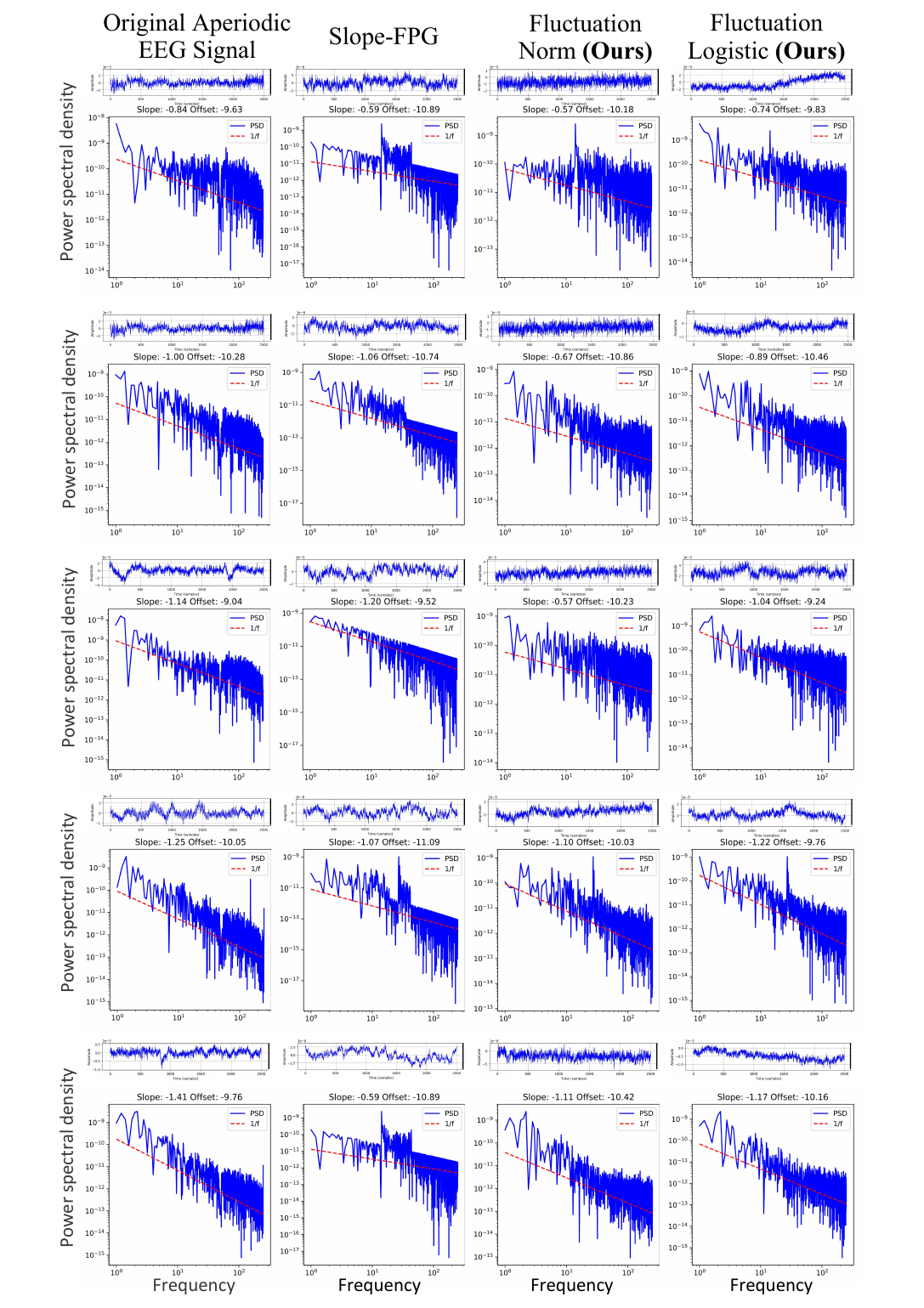}
  \caption{\textbf{Additional examples of aperiodic EEG signal generation using different methods, spanning a range of 1/f slopes.}  1/f slope fitting is acquired using FOOOF~\citep{FOOOF}. Our fluctuation modeling method is capable of generating a wide range of 1/f slopes, while preserving a slope similar to that of the original signal.}
  \label{fig:appendix_signals}
\end{figure}

\clearpage

\end{document}